\documentclass{article}      %2.8.00 f
\usepackage[dvips]{graphicx,color} \usepackage{epsfig}
\DeclareGraphicsExtensions{.ep.gz,.eps,.ps,lps.gz}
\newcommand{\omm}{\Omega_{m0}}
\newcommand{\omlam}{\Omega_{\Lambda}}
\newcommand{\omq}{\Omega_{Q0}}

\newcommand{\bc}{\begin{center}}
\newcommand{\ec}{\end{center}}
\begin{document}         %follows before \title
\title{\bf OBSERVED SMOOTH ENERGY IS ANTHROPICALLY EVEN MORE LIKELY AS
  QUINTESSENCE THAN AS COSMOLOGICAL CONSTANT}

\author{Sidney Bludman\\Deutsches Elektronen-Synchrotron DESY, Hamburg\\
  University of Pennsylvania, Philadelphia \thanks{Supported in
    part by Department of Energy grant DE-FGO2-95ER40893}}

\maketitle

\abstract{For a universe presently dominated by static or dynamic
  vacuum energy, cosmological constant (LCDM) or quintessence (QCDM),
  we calculate the asymptotic collapsed mass fraction as function of
  the present ratio of vacuum energy to clustered mass, $\omq/\omm$.
  Identifying these collapsed fractions as anthropic probabilities, we
  find the present ratio $\omq/\omm \sim 2$ to be reasonably likely in LCDM,
  and very likely in QCDM.}
\section {A Cosmological Constant or Quintessence?}

Absent a symmetry principle protecting its value, no theoretical
reason for making the cosmological constant zero or small has been
found.  Inflation makes the universe flat, so that, at present, the
vacuum or smooth energy density $\Omega_{Q0}=1-\omm < 1$, is $10^{120}$
times smaller than would be expected on current particle theories.  To
explain this small but non-vanishing present value, a dynamic vacuum energy,
quintessence, has been invoked, which obeys the equation
of state $w_Q \equiv P/\rho <0$.  (The limiting case, $w_Q=-1$, a static
vacuum energy or Cosmological Constant, is homogeneous on all scales.)

Accepting this small but non-vanishing value for static or dynamic
vacuum energy, the {\em Cosmic Coincidence} problem now becomes
pressing: Why do we live when the clustered matter density
$\Omega(a)$, which is diluting as $a^{-3}$ with cosmic scale $a$, is
just now
comparable to the static vacuum energy or present value of the smooth
energy:
   $$u_0^3 \equiv \Omega_{Q0}/\omm \sim 2  . $$

The observational evidence\cite{WCOS}
is for a flat, low-density universe:\\
 
(1) $\Omega_m+\Omega_Q=1 \pm 0.2$ \\
(Location of first Doppler peak in the CBR anisotroy at $l \sim 200$);\\
(2) $\omm=0.3 \pm 0.05$. (Slow evolution
of rich clusters, mass power spectrum, CBR anisotropy, cosmic flows);\\
(3) $\omq=1-\omm \sim 2/3$ (curvature in SNIa Hubble diagram, dynamic
age,height of first Doppler peak, cluster evolution).\\
Of these, the SNIa evidence is most subject to systematic errors due
to precursor intrinsic evolution and the possibilty of grey dust extinction.

The combined data implies a flat, low-density universe with $\omm \sim
1/3$, with negative pressure $ -1 \leq w_Q \leq -1/2 $.  In this
paper, we use the evolution of large-scale structure to distinguish
the two limiting cases:\\ \\
LCDM: Cosmological constant: $w_Q=-1, \quad n_Q\equiv 3(1+w_Q)=0 \quad \omlam =2/3 $ \\
QCDM: Quintessence: $ w_Q =-1/2, n_Q=3/2,\quad \Omega_{Q0}=1/3$ .

\section {Evolution of a Low Density Flat Universe}

The Friedmann equation in a flat universe with clustered matter and
smooth energy density is
$$
H^2(x) \equiv (\dot{a}/a)^2=(8 \pi G/3)(\rho_m+\rho_Q), $$
or, in units of $\rho_{cr}(x)=3H^2(x)/8\pi G$,
$1=\Omega_m(x)+\Omega_Q(x),$ where the reciprocal scale factor $x \equiv
a_0/a \equiv 1+z \rightarrow \infty$ in the far past, $\rightarrow 0$
in the far future.

With the EOS $w \equiv P/\rho$, different kinds of energy density dilute at
different rates $\rho \sim a^{-n},~n \equiv 3(1+w)$, and contribute to
the deceleration at different rates $(1+3w)/2$ shown in the table:\\ 
%%%%%%%%%%%%%%%%%%%%%%%%%%%%%%%%%%%%%%%%%%%%%%%%%%%%%%%%%%%%%%%%%%%%%%%%%%%%%%%
   \begin{table}[h]{\bfseries Energy Dilution for Various Equations of
       State}\\[1ex]
   \centering
   \begin{tabular*}{115mm}{@{\extracolsep{\fill}}l|ccc@{}}  \hline
   
   {\em substance}              &{\em w}       &{\em n}       &{\em (1+3w)/2}   \\  \hline
   radiation      &  1/3   &  4     &1    \\
   NR matter      &   0    &  3     &1/2      \\
   quintessence   & -1/2   &  3/2   &-1/4   \\
   cosmolconst    & -1     &  0     &-1     \\
   \hline  
   \end{tabular*}\\[0.5ex]
   \end{table}\\
%%%%%%%%%%%%%%%%%%%%%%%%%%%%%%%%%%%%%%%%%%%%%%%%%%%%%%%%%%%%%%%%%%
The expansion rate in present Hubble units is 
$$ E(x) \equiv H(x)/H_0=(\omm x^3+ (1-\omm) x^n_Q)^{1/2}. $$
The Friedmann equation has an unstable fixed point
in the far past and a stable attractor in the far future.  (Note the
tacit application of the anthropic principle: Why does our universe
expand, rather than contract?)

The second Friedmann equation is $-\ddot{a} a/{\dot{a}^2}=(1+3w_Q
\Omega_Q)/2 $. The ratio of smooth energy to matter energy,
$\Omega_Q/\Omega_m \equiv u^3=u_0^3x^{3w_Q}$, where $ \Omega_{Q0} /
\omm \equiv u_0^3 \sim 2$ is the present ratio. 
As shown by the inflection points in the middle curves of the figure,
for fixed $\Omega_{Q0}/\Omega_{m0}$, QCDM (upper middle curve) expands
faster than LCDM (lower middle curve), but begins accelerating only
at the present epoch.  The top and bottom curves refer respectively to
a De Sitter universe ($\Omega_m=0$), which is always accelerating, and
an SCDM universe ($\Omega_m=1$), which is always decelerating.
\nopagebreak
%%%%%%%%%%%%%%%%%%%%%%%%%%%%%%%%%%%%%%%%%%%%%%%%%%%%%%%%%%%%%%%%%%%%%%%
\begin{figure}[b]
\begin{center} 
\epsfig{file=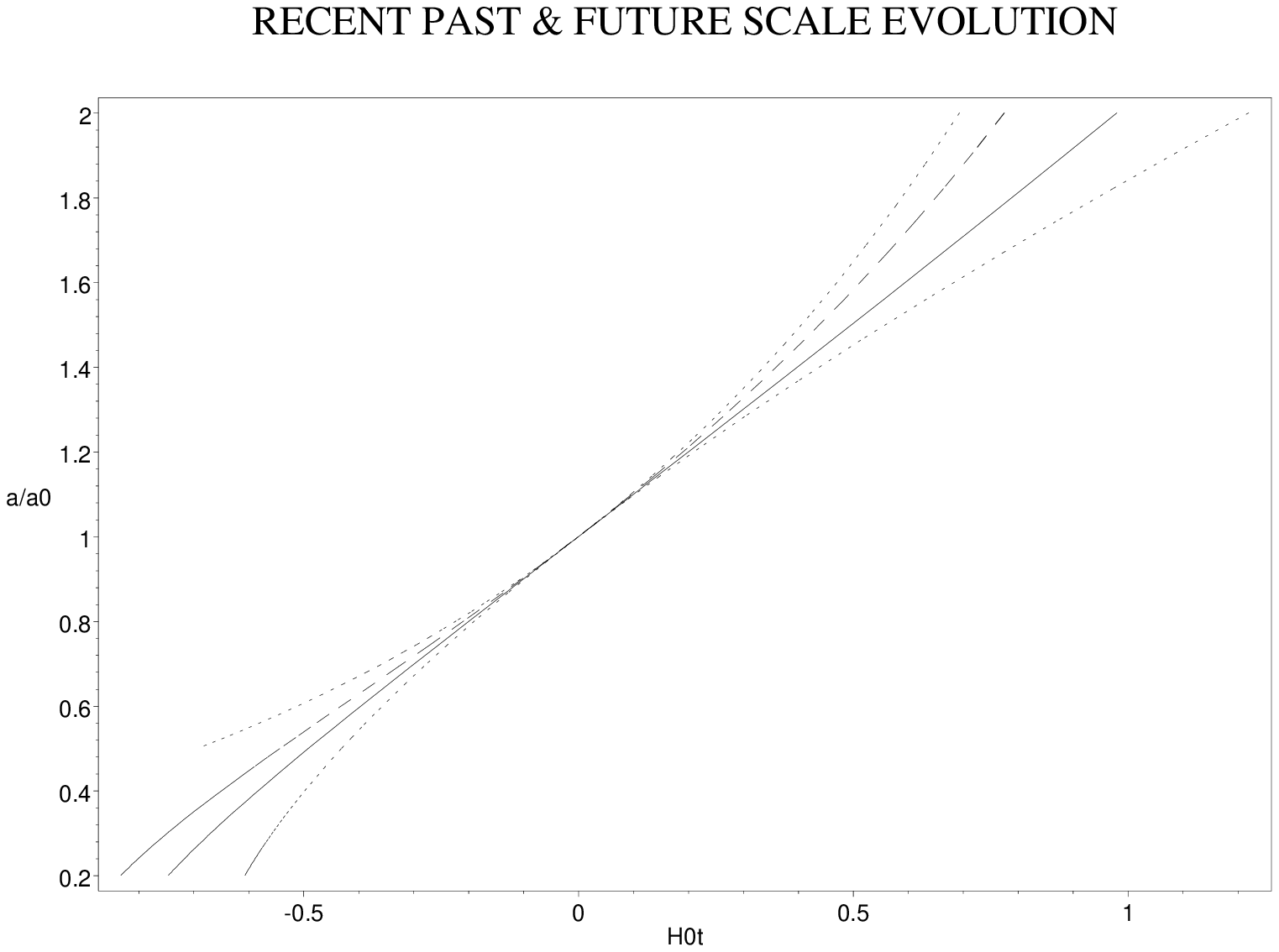,width=12cm,height=6.5cm} 
\end{center}      \end{figure}
%\epsfig{} could include adding bounding box to .ps file, but I have already
%written bounding box onto top of .eps. 
%%%%%%%%%%%%%%%%%%%%%%%%%%%%%%%%%%%%%%%%%%%%%%%%%%%%%%%%%%%%%%

\pagebreak
As summarized in the table below, quintessence dominance begins 3.6
 Gyr earlier and more gradually than cosmological constant dominance.
 (In this table, the deceleration $q(x) \equiv -\ddot{a}/aH_0^2$ is
 measured in {\em present} Hubble units.)  The recent lookback time
$$H_0t_L(z)=z-(1+q_0)z^2+...,\quad z<1 ,$$
where $q_0=0$ for QCDM and $=- 1/2$ for LCDM.

%%%%%%%%%%%%%%%%%%%%%%%%%%%%%%%%%%%%%%%%%%%%%%%%%%%%%%%%%%%%%%%%%%%%%%%%%%%
   \begin{table}[h]{\bfseries  Comparative Evolution of LCDM and  QCDM}\\[1ex] 
   \centering
   \noindent
   \begin{tabular*}{125mm}{@{\extracolsep{\fill}}l|cc@{}}  \hline
   {\em event}                 &  {\em LCDM}      &  {\em QCDM} \\
   \hline \hline
   {\bf Onset of Vacuum Dominance}&               &                \\
   reciprocal scale x*=$a_0/a=1+z$                &$u_0$=1.260           &$u_0^2$=1.587     \\
   age   t(x*) ($H_0^{-1}$)    &0.720             &0.478           \\
         $h_{65}^{-1}$Gyr      &10.8              &7.2            \\ \hline
   horizon(x*) ($cH_0^{-1})$   &2.39              &1.58           \\
         $h_{65}^{-1}$Gpc      &11.0              &7.24            \\ \hline
   deceleration q(x*) at freeze-out               &-0.333            &0.333
   \\  \hline \hline
   {\bf Present Epoch}         &                  &              \\
   age t0 ($H_0^{-1}$)         &0.936             &0.845          \\
         $h_{65}^{-1}$Gyr      &14.0              &12.7           \\ \hline
   horizon                     &3.26              &2.96          \\
          $h_{65}^{-1}$Gpc     &15.0              &13.6            \\ \hline
   present deceleration $q_0$  &-0.500            &0              \\ \hline

   \end{tabular*}
   \end{table}
%%%%%%%%%%%%%%%%%%%%%%%%%%%%%%%%%%%%%%%%%%%%%%%%%%%%%%%%%%%%%%%%%%%%%%%%%%%
 
 The density ratio $u^3(a)\equiv \Omega_Q/\Omega_m=u_0^3 x^{3 w_Q}$,
 increases as the matter density decreases.  The
 matter-smooth energy transition $\Omega_Q/\Omega_m=1$ took place only
 recently at $x*^{-w_Q}=u_0$ or at $x*=1+z*=u_0^2=1.5874$ for QCDM
 and, even later, at $x*=1+z*=u_0=1.260$ for LCDM. Because, for the
 same value of $u_0$, a matter-QCDM freeze-out would take place
 earlier and more slowly than a matter-LCDM freeze-out, it imposes a
 stronger constraint on structure evolution.  To permit evolution to the same
 present structure, QCDM would require a smaller value of $\omq/\omm$ than
 does LCDM.

\section{Growth of Large Scale Structure}

The background density for large-scale structure formation
  is overwhelmingly Cold Dark Matter (CDM), consisting of clustered
  matter $\Omega_m$ and smooth energy or quintessence $\Omega_Q$.
  Baryons, contributing only a fraction to $\Omega_m$, collapse after
  the CDM and, particularly in small systems, produce the large
  overdensities that we see. 
  
  Structure formation begins and ends with matter dominance, and is
  characterized by two scales: The horizon scale at the first
  cross-over, from radiation to matter dominance, determines the power
  spectrum $P(k,a)$, which is presently characterized by a scale
  factor $\Gamma=\omm h =0.25 \pm 0.05$.  The horizon scale at the
  second cross-over, from matter to smooth energy, determines a second
  scale factor, which for quintessence, is $\Gamma_Q$ at $\sim
  130~Mpc$, the scale of voids, superclusters.  A cosmological
  constant is smooth at all scales.

  Quasars formed as far back as $z \sim 5$, galaxies at $z \geq 6.7$,
  ionizing sources at $z=(10-30)$.  The formation of {\em any} such
  structures, already sets an upper bound $x*<30$ or $(\omlam /
  \omm)<1000, \omq<30$, for {\em any} structure to have formed.  A
  much stronger upper bound, $u_0<5$, is set by when {\em typical}
  galaxies form i.e. by using the observed LSS, not to fix $\omlam$ or
  $u_0^3$, but to estimate the probability of our observing this ratio
  $\omq/\omm$ at the present epoch.
  
  For LCDM, Martel {\it et al} \cite{MSW} and Garriga {\it et
    al} \cite{MS} calculate the asymptotic mass fraction that
  ultimately collapses into galaxies to be
$$f_{c,\infty}=\mbox{erfc}(\beta^{1/2}),$$
remarkably a broad function of only $\beta
\equiv \delta_{i,c}^2/2(\sigma_i)^2$, where
$\sigma_i^2=(1.7-2.3)/(1+z_i)$ is the variance of the mass power
spectrum and $\delta_{i,c}$
is the minimum density contrast which will
make an ultimately bound perturbation.  This minimum density contrast
grows with scale factor $a$, and is approximately unity at
recombination.  Thus, except for a numerical
factor of order unity \cite{MS}, $\delta_{i,c} \sim x*/(1+z_i)$, 
the freeze-out projected back to recombination.
Both numerator and denominator
in $\beta$ refer to the time of recombination, but this initial time
or red-shift cancels out in the quotient.

\section{$\Omega_Q \sim \Omega_m$ is 
Quite Likely for Our Universe}

For a cosmological constant, an anthropic argument has
already been given \cite{E,V,MSW,GLV}, assuming a universe of subuniverses
with all possible values for the vacuum energy $\rho_V$ or $\omlam$.
In each of these subuniverses, the probability
for habitable galaxies to have emerged before the present epoch, is a
function of $\omlam$ or the present ratio $\omlam/\omm$ 
$$\mathcal{P}(\rho_V) \propto (\mbox{prior distribution in} ~\rho_V)
\times(\mbox{asymptotic mass fraction} ~f_{c,\infty}). $$
MSW,
assuming nothing about initial conditions, assume a prior flat in
$\omlam$.  GLV argue that the prior should be determined by a theory
of initial conditions and is {\em not} flat for most theories.
 
Following MSW, we assume a flat prior, so that the differential
probability $\mathcal{P}$ for our being here to observe a value
$\rho_V$ in our universe is simply proportional to the asymptotic
collapsed mass fraction for this $\rho_V$.  For LCDM,
$$\delta_{i,c}=1.1337 u_0/(1+z_i) , \quad \quad 1.1337=(27/2)^{2/3}/5.
$$
As function of the ratio $\omlam/\omm=u_0^3$, the LCDM probability
distribution has a broad peak about $u_0^3 \approx 12-30$.  The value
observed in our universe $u_0^3 \approx 2$ has reasonable probability
$4-10\%$.

This argument \cite{MSW,GLV} for LCDM ($w_Q=-1$) is easily extended to
QCDM ($w_Q=-1/2$).  The variance of the power spectrum, $\sigma^2$, is
insensitive to $w_Q$ for $w_Q<-1/3$ \cite{WS}. For $w_Q=-1/2$, the numerical
factor in $\delta_{i,c}$ is the same as for $w_Q=-1$, but $x*=u_0^2$ in
place of $u_0$, 
so that $\delta_{i,c}=1.1337
u_0^2/(1+z_i)$.  Thus $\beta_{QCDM}(u_0)=\beta_{LCDM}(\sqrt{u_0})$, so
that the QCDM probability distribution now peaks at $u_0^3 \approx
3.5-5.5$.  With QCDM, the probability for observing $u_0^3 \approx 2$
is now increased to about 50\%.

\end{document}